\documentclass[12pt]{article}
\usepackage{color,epsfig,graphicx,latexsym,subfigure,url}

\title{Hinged Dissection of Polyominoes and Polyforms}
\author{%
  Erik D. Demaine%
    \thanks{MIT Laboratory for Computer Science, 200 Technology Square,
      Cambridge, MA 02139, USA,
      \{\texttt{edemaine}, \texttt{mdemaine}\}\texttt{@mit.edu}}
  \and
  Martin L. Demaine\footnotemark[1]
  \andlinebreak
  David Eppstein%
    \thanks{Department of Information and Computer Science, University of
      California, Irvine, CA 92697, USA, email:
      \texttt{eppstein@ics.uci.edu}.}
  \and
  Greg N. Frederickson%
    \thanks{Department of Computer Sciences, Purdue University,
      West Lafayette, IN 47907-1398, USA, email:
      \texttt{gnf@cs.purdue.edu}.}
  \and
  Erich Friedman%
    \thanks{Mathematics Department, Stetson University, DeLand, FL 32720,
      USA, email: \texttt{efriedma@stetson.edu}.}
}
\date{}

\advance \topmargin by -\headheight
\advance \topmargin by -\headsep
\evensidemargin \oddsidemargin
\let\latexcite=\cite
\def\cite{\nolinebreak\latexcite}
\let\latexref=\ref
\def\ref{\nolinebreak\latexref}

{\makeatletter
 \gdef\xxxmark{%
   \expandafter\ifx\csname @mpargs\endcsname\relax 
     \expandafter\ifx\csname @captype\endcsname\relax 
       \marginpar{xxx}
     \else
       xxx 
     \fi
   \else
     xxx 
   \fi}
 \gdef\xxx{\@ifnextchar[\xxx@lab\xxx@nolab}
 \long\gdef\xxx@lab[#1]#2{{\bf [\xxxmark #2 ---{\sc #1}]}}
 \long\gdef\xxx@nolab#1{{\bf [\xxxmark #1]}}
}

\def\andlinebreak{\end{tabular}\linebreak\begin{tabular}[t]{c}}

{\makeatletter \gdef\fps@figure{!htbp}}

\newtheorem{theorem}{Theorem}
\newtheorem{lemma}[theorem]{Lemma}
\newtheorem{proposition}[theorem]{Proposition}
\newtheorem{corollary}[theorem]{Corollary}

\makeatletter
\def\GrabProofArgument[#1]{ (#1): \egroup\ignorespaces}
\def\proof{\noindent\textbf\bgroup Proof%
           \@ifnextchar[{\GrabProofArgument}{: \egroup\ignorespaces}}

\makeatother

\begin{document}
\maketitle


\begin{abstract}
A hinged dissection of a set of polygons $S$ is a collection of polygonal
pieces hinged together at vertices that can be rotated into any member of $S$.
We present a hinged dissection of all edge-to-edge gluings of $n$ congruent
copies of a polygon $P$
that join corresponding edges of $P$.  This construction uses $kn$
pieces, where $k$ is the number of vertices of $P$.  When $P$ is a regular
polygon, we show how to reduce the number of pieces to $\lceil k/2 \rceil
(n-1)$.  In particular, we consider polyominoes (made up of unit squares),
polyiamonds (made up of equilateral triangles), and polyhexes (made up of
regular hexagons).  We also give a hinged dissection of all polyabolos (made up
of right isosceles triangles), which do not fall under the general result
mentioned above.  Finally, we show that if $P$ can be hinged into $Q$, then any
edge-to-edge gluing of $n$ congruent copies of $P$ can be hinged into any
edge-to-edge gluing of $n$ congruent copies of $Q$.
\end{abstract}

\section{Introduction}

A {\it geometric dissection} \cite{Frederickson-1997, Lindgren-1972} is a
cutting of a polygon into pieces that can be re-arranged to form another
polygon.
It is well known, for example, that any polygon can be
dissected into any other polygon with the same area
\cite{Boltianskii-1978-2Dpart, Frederickson-1997, Lowry-1814}, but the bound on
the number of pieces is quite weak.  The main problem, then, is to find a
dissection with the fewest possible number of pieces.
%
Dissections have begun to be studied more formally than in their recreational
past.  For example, Kranakis, Krizanc, and Urrutia
\cite{Kranakis-Krizanc-Urrutia-2000} study the asymptotic number of pieces
required to dissect a regular $m$-gon into a regular $n$-gon.  Czyzowicz,
Kranakis, and Urrutia \cite{Czyzowicz-Kranakis-Urrutia-1999} consider the
number of pieces to dissect a rational rectangle into a square using ``glass
cuts.''  An earlier paper by Cohn \cite{Cohn-1975} studies the number of pieces
to dissect a given triangle into a square.

\begin{figure}
\centerline{%
  $\vcenter{\hbox{\includegraphics[scale=0.25]{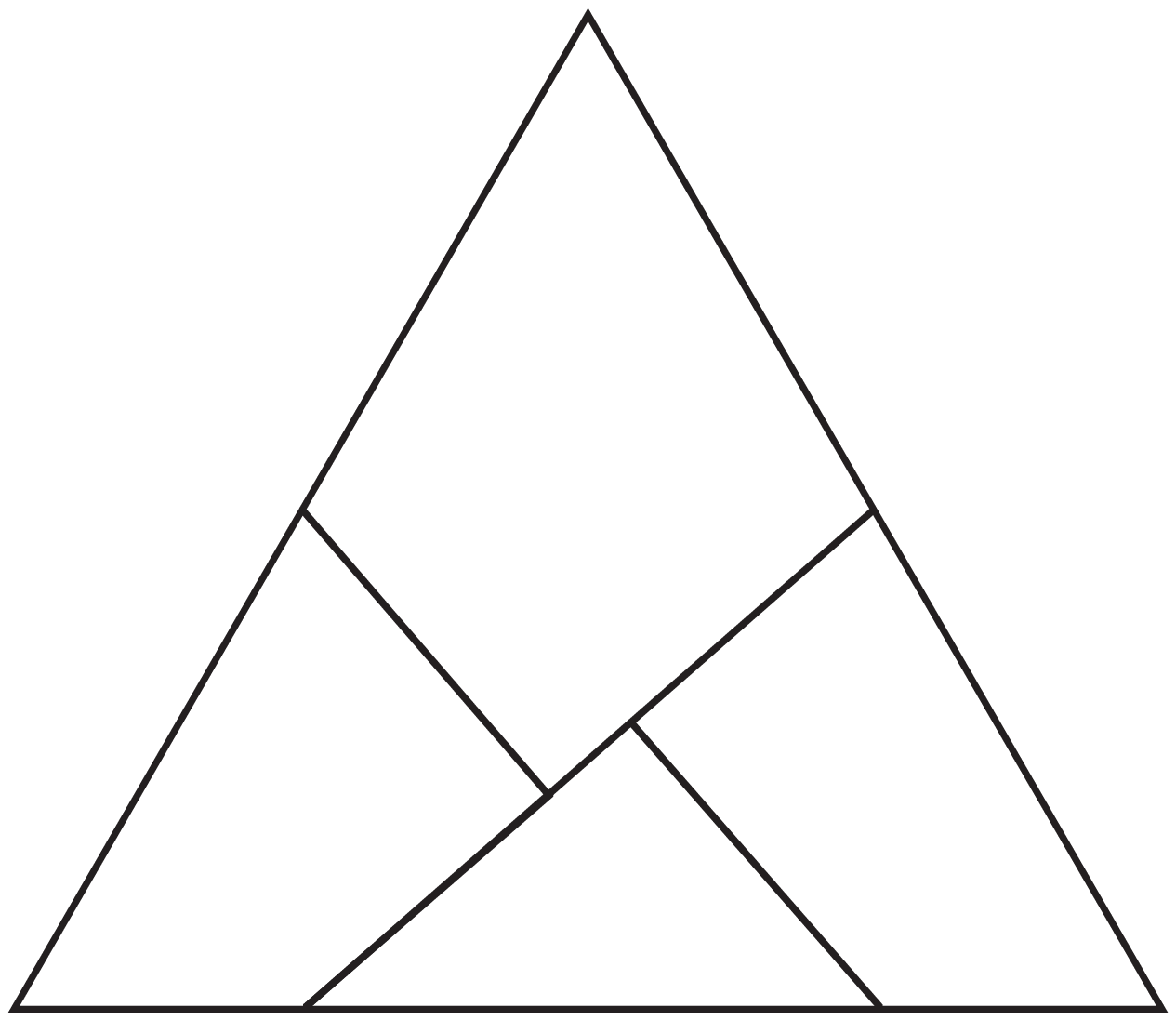}}}$
  \qquad
  $\vcenter{\hbox{\includegraphics[scale=0.25]{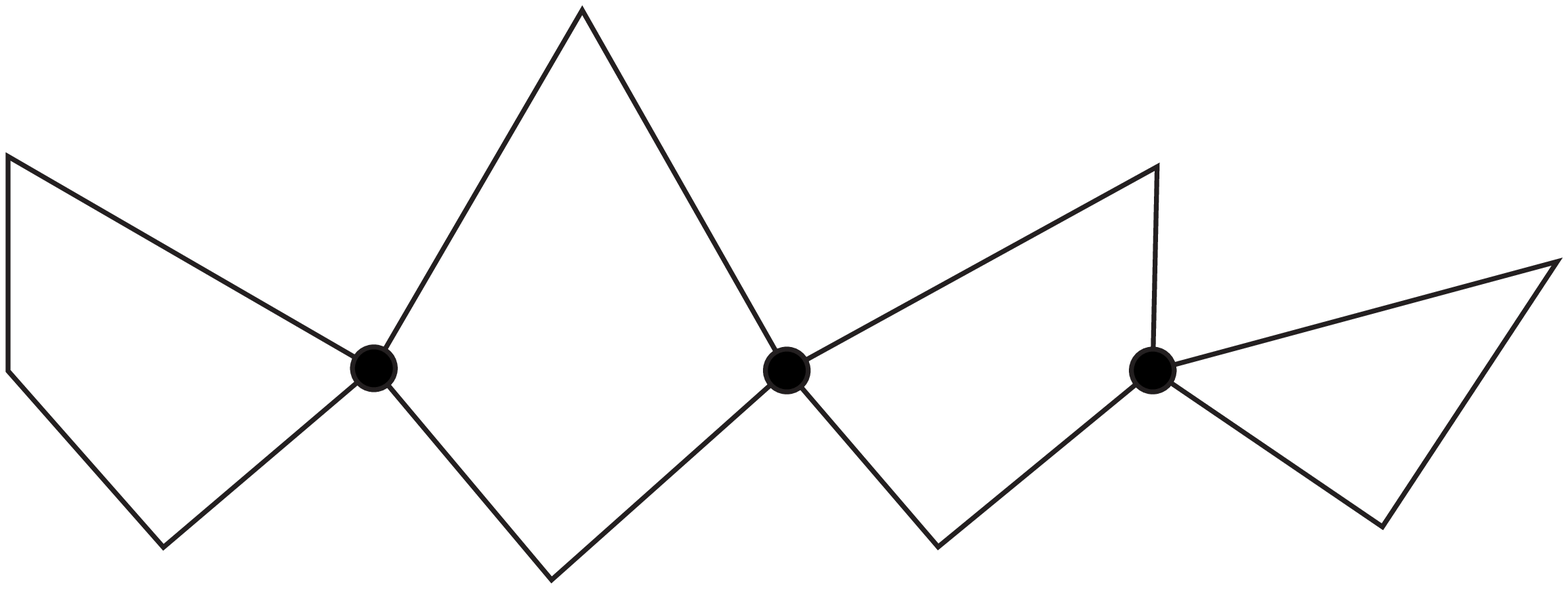}}}$
  \qquad\quad
  $\vcenter{\hbox{\includegraphics[scale=0.25]{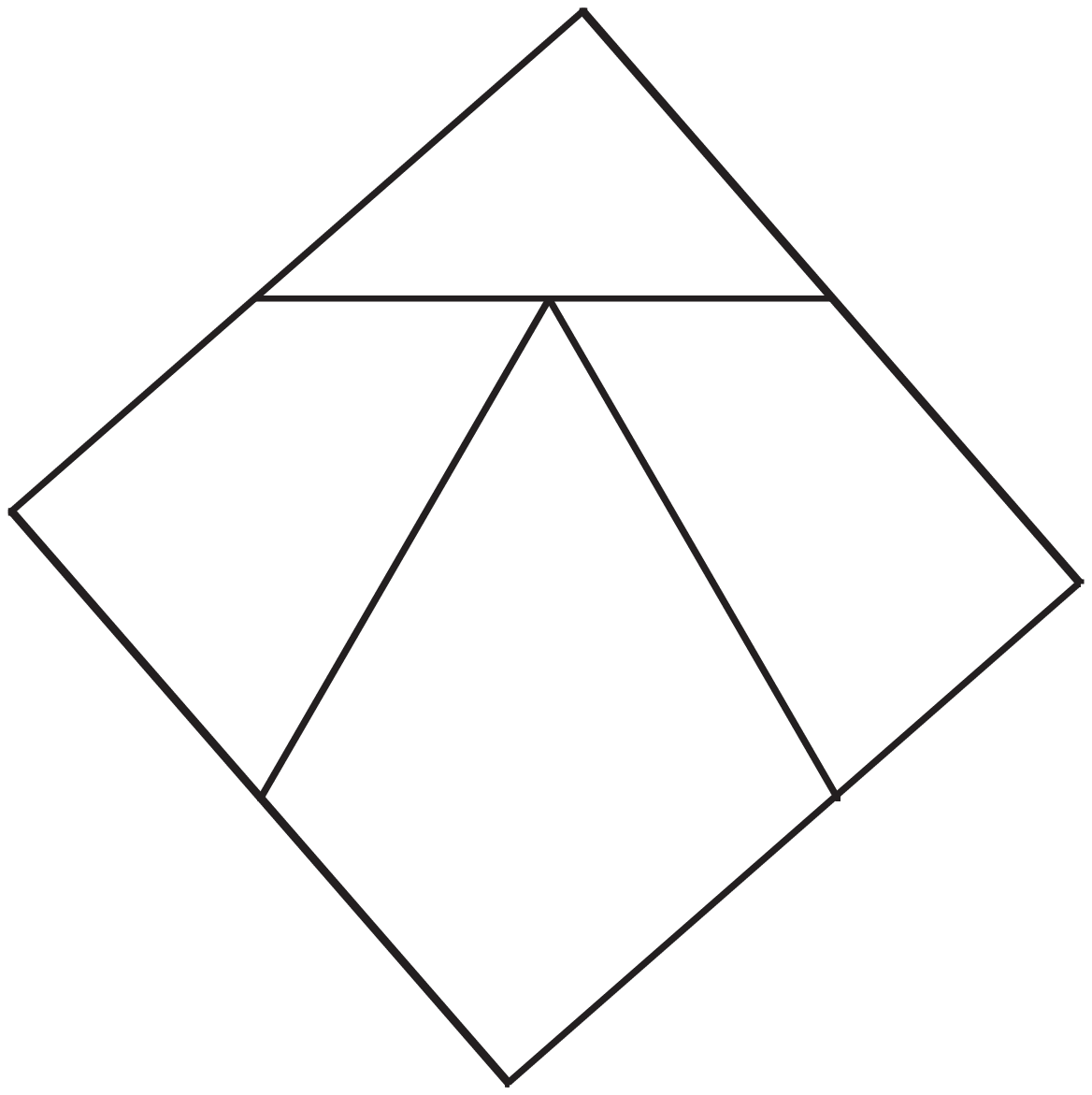}}}$
}
\caption{\label{Dudeney}
  Hinged dissection of an equilateral triangle into a square.}
\end{figure}

An intriguing subclass of dissections are \emph{hinged dissections}
\cite{Frederickson-2002, Frederickson-1997}.
Instead of allowing the pieces to be re-arranged arbitrarily, suppose that the
pieces are hinged together at their vertices, and we require pieces to remain
attached at these hinges as they are re-arranged.
Figure \ref{Dudeney} shows
the classic hinged dissection of an equilateral triangle into a square.  This
dissection is described by Dudeney \cite{Dudeney-1902-hinged}, but may have
been discovered by C. W. McElroy; see \cite{Frederickson-2002},
\cite[p.~136]{Frederickson-1997}.

\begin{figure}
\centerline{\input{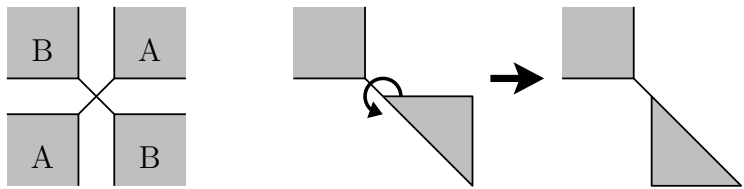}}
\caption{\label{rules}
  Forbidden features of hinged dissections:
  no crossing hinges (left) and no hinge twisting (right).}
\end{figure}

For our purposes, we allow the pieces in a hinged dissection to overlap
as the hinges rotate, but are interested in final configurations
at which pieces do not overlap.
We do not allow multiple hinges at a common vertex to cross each
other, nor for hinges to ``twist'' and flip pieces over;
see Figure \ref{rules}.%
\footnote{Frederickson \cite{Frederickson-2002} distinguishes different types
  of hinged dissections; this type is called \emph{swing-hinged}
  (no twisting) and \emph{wobbly hinged} (allow overlap during rotation).}
Figures \ref{Dudeney} and \ref{rules} illustrate our two drawing styles for
hinged dissections: ``geometrically exact'' with dots for hinges and unshaded
pieces, and ``exaggerated'' with segments for hinges and shaded pieces.
In fact, hinges have zero length.

A natural question about hinged dissections is the following: can any polygon
be hinge-dissected into any other polygon with the same area?  This question is
open and seems quite difficult.  The main impediment to applying the same
techniques as normal dissection is that hinged dissections are not obviously
transitive: if $A$ can be hinge-dissected to $B$, and $B$ can be
hinge-dissected to $C$, then it is not clear how to combine the two dissections
into one from $A$ to $C$.  Of course, this transitivity property holds for
normal dissections.

The possibility of an affirmative answer to this question is supported by
the many examples of hinged dissections that have been discovered.
Frederickson \cite{Frederickson-2002} has developed several
techniques for constructing hinged dissections, and has applied them to
design hundreds of examples.
Akiyama and Nakamura \cite{Akiyama-Nakamura-1998} have demonstrated some hinged
dissections under a restrictive model of hinging, designed to match the
dissection in Figure \ref{Dudeney}.  For example, they show that it is possible
to hinge-dissect any convex quadrangle into some parallelogram; in general,
their work only deals with polygons having a constant number of vertices.
Eppstein \cite{Eppstein-2001-mirror-dissection} gives a general method for
hinge-dissecting any $n$-vertex polygon into its mirror image
using $O(n)$ pieces.  His method also reduces the general hinged-dissection
problem to determining whether there is a hinged dissection between every pair
of equal-area triangles satisfying a few simple extra properties.

In this paper, we explore hinged dissections of a class of polygons formed
by gluing together several nonoverlapping equal-size regular $k$-gons
along touching pairs of edges, for a fixed $k$.
We call such a polygon a \emph{poly-$k$-regular} or
\emph{polyregular} for short.  The polygon need not be simply connected; we
allow it to have holes.  An \emph{$n \times k$-regular} is a poly-$k$-regular
made of $n$ regular $k$-gons.
Poly-$k$-regulars include the well-studied \emph{polyominoes} ($k = 4$),
\emph{polyiamonds} ($k = 3$), and \emph{polyhexes} ($k = 6$) \cite{Golomb-1994,
Klarner-1997, Martin-1991}.
Polyominoes are of particular interest to
computational geometers, because they include orthogonal polygons whose
vertices have rational coordinates.

This paper proves that not only can any $n \times k$-regular be hinge-dissected
into any other $n \times k$-regular, but furthermore that there is a single
hinged dissection that can be rotated into all $n \times k$-regulars, for fixed
$n$ and $k$.  This includes both reflected copies of each polyregular.
Section \ref{Polyregulars} describes two methods for solving this problem,
the more efficient of which uses $\lceil k/2\rceil (n-1)$ pieces.
The more-efficient method combines a simpler method, which uses $k(n-1)$
pieces, and an efficient method for the special case of polyominoes,
described in Section \ref{Polyominoes} where we also give some lower bounds.

Next, in Section \ref{Polyabolos} we consider another kind of ``polyform.''  A
\emph{polyabolo} is a connected edge-to-edge gluing of nonoverlapping
equal-size right isosceles triangles.  In particular, every $n$-omino is a
$2n$-abolo, as well as a $4n$-abolo.  We prove that there is a $4n$-piece
hinged dissection that can be rotated into any $n$-abolo for fixed $n$.

In Section \ref{Other Polyforms}, we show an analogous result for a general
kind of polyform, which allows us to take certain edge-to-edge gluings of
copies of a general polygon.  This result is a generalization of polyregulars,
although it uses more pieces.  It does not, however, include the polyabolo
result, because of some restrictions placed on how the copies of the polygon
can be joined.

Section \ref{Polyforms of Different Types} shows there are hinged dissections
that can rotate into polyforms made up of different kinds of polygons.
Specifically, it finds efficient hinged dissections that rotate into
(a) all polyiamonds and all polyominoes,
(b) all polyominoes and all polyhexes, and
(c) all polyiamonds and all polyhexes.
On the way, we show that any dissection in which the pieces
are hinged together according to some graph, such as a path, can be turned into
a dissection in which the pieces are hinged together in a cycle.



A preliminary version of this work appeared in
\cite{Demaine-Demaine-Eppstein-Friedman-1999}.

\section{Basic Structure of Polyforms}
\label{Basic Structure of Polyforms}

We begin with some definitions and basic results about the structure of
polyforms.  As we understand it, the term ``polyform'' is not normally used in
a formal sense, but rather as a figurative term for objects like polyominoes,
polyiamonds, polyhexes, and polyabolos.  However, in this paper, we find it
useful to use a common term to specify all of these objects collectively, and
``polyform'' seems a natural term for this purpose.

Specifically, we define a (planar) \emph{polyform} to be a finite collection of
copies of a common polygon $P$ such that the interior of their union
is connected, and the intersection of two copies is either empty, a common
vertex, or a common edge.  An $n$-form is a polyform made of $n$ copies of $P$.
We call $P$ the \emph{type} of the polyform.
Polyforms are considered equivalent modulo rigid motions (translations
and rotations), but not reflections.

The \emph{graph} of a polyform is defined as follows.  Create a vertex for
each polygon in the collection, and connect two vertices precisely if the
corresponding polygons share an edge.  Because every connected graph has a
vertex whose removal leaves the graph connected, we have the following
immediate consequence.

\begin{proposition} \label{induction} 
Every $n$-form has a polygon whose removal results in a (connected)
$(n-1)$-form of the same type.
\end{proposition}

This simple result is useful for performing induction on the number of polygons
in a polyform.  More precisely, if we view the decomposition in the reverse
direction (adding polygons instead of removing them), then this lemma says that
any polyform can be built up by a sequence of additions such that any
intermediate form is also connected.  To construct a hinged dissection, we will
repeatedly hinge a new polygon onto the previously constructed polyform.

In the next lemma, the first sentence restates Proposition \ref{induction} in
the context of adding polygons instead of removing them.  In addition, the
second sentence provides some additional constraints on the addition process,
which will be important for optimizing some of our dissections.

\begin{lemma} \label{two-end induction}
Any polyform of type $P$ can be built up by a sequence of \emph{gluings} in
which a new copy of $P$ is placed against an edge of an already placed copy
of $P$ called the \emph{parent}.
(As a special case, the first copy of $P$ can be placed arbitrarily.)
Furthermore, this gluing sequence can be chosen so that only one copy
of $P$ is glued with the first copy of $P$ as the parent.
\end{lemma}

\begin{proof}
Pick any spanning tree $T$ of the graph of an $n$-form.  Let $P_1$ be some leaf
of $T$.  Let $P_2$ denote the unique vertex incident to $P_1$ in $T$, and glue
it to $P_1$.  Now perform a depth-first traversal of $T$ rooted at $P_2$, and
label newly visited vertices as $P_3, P_4, \dots, P_n$, each gluing to its
parent.  The result is a gluing sequence for the polyform, such that only one
polygon ($P_2$) glues to $P_1$.
\end{proof}

Our constructions and proofs of correctness for hinged dissections of $n$-forms
will follow a common outline.
The construction is simple: we describe a (often cyclicly) hinged dissection
parameterized by $n$; more precisely we define a function from the positive
integers to hinged dissections, call it $H(n)$.  Now we need to prove that, for
any $n \geq 1$, $H(n)$ can be rotated into all $n$-forms.  This will be done
using the ideas of Lemma \ref{two-end induction}.  First we show how to
construct a single polygon ($P_1$); that is, we show that $H(1)$ can be rotated
into $P$.  Second we show how to add each polygon $P_n$ onto a rotation of
$H(n-1)$ into an arbitrary $(n{-}1)$-form $F_{n-1}$, so that in the end we have
a rotation of $H(n)$ into a desired $n$-form.
The key is that no matter where we attach $P_n$ to $F_{n-1}$, we obtain a
rotation of exactly the same hinged dissection, $H(n)$.  This means that the
same $H(n)$ can be rotated into all of these configurations---all $n$-forms.

This technique will be used repeatedly, so the following lemma specifies it
formally.  It also generalizes to allow starting with $c$-forms for a constant
$c \geq 1$ instead of just a single copy of $P$.  We will often use $c=2$ to
optimize some of our dissections, although we will never use higher values of
$c$.

\begin{lemma} \label{build up}
For any constant $c \geq 1$, a parameterized hinged dissection $H(n)$ rotates
into all $n$-forms of type $P$, for every $n \geq c$, if
  \begin{enumerate}
  \item $H(c)$ rotates into any $c$-form of type $P$, and
  \item for every $n > c$, given any rotation of $H(n-1)$ into an
        $(n{-}1)$-form $F$, and given any edge $e$ of $F$,
        $H(1)$ can be rotated into $P$, placed
        next to $e$, and hinges can be added and removed between
        touching vertices (that is, the two hinged dissections can be split
        up and spliced back together) such that the resulting hinged dissection
        is $H(n)$.  This process is called \emph{attaching} a copy of $P$
        to $F$.
  \end{enumerate}
Furthermore, if $c \geq 2$, some copy of $P$ in the initial $c$-form (from
Condition 1) will never have a copy of $P$ attached to it, and in this sense is
called \emph{slippery}.
\end{lemma}

\begin{proof}
Consider any $n$-form $F$ of type $P$ for $n \geq c$.  We will prove that
$H(n)$ rotates into $F$, and hence $H(n)$ rotates into all $n$-forms of type $P$.
Consider a gluing sequence for $F$ from Lemma \ref{two-end induction}.  The
union of the first $c$ copies of $P$ is some $c$-form $C$ of type $P$.
By Condition 1, $H(c)$ can be rotated into $C$.
Now attach the remaining copies of $P$ one by one in the order specified by the
gluing sequence.  At each step, if we have a rotation of $H(k)$ into the first
$k$ copies of $P$, then attaching the next gives us a rotation of $H(k+1)$.
By induction, we reach a rotation of $H(n)$ into the desired $n$-form $F$.

Now $C$ contains the first copy of $P$ in the gluing sequence, call it $P_1$.
The gluing sequence from Lemma \ref{two-end induction} glues only one copy of
$P$ to $P_1$.  Provided $c \geq 2$, $C$ has already glued a copy of $P$ to
$P_1$, and hence no other copies of $P$ are glued to $P_1$.  The above
construction makes precisely one attachment corresponding to each gluing other
than the first $c-1$ gluings; thus, no copies of $P$ are ever attached to
$P_1$.
\end{proof}

\section{Polyominoes}
\label{Polyominoes}

Let us start with the special case of polyominoes.  This serves as a nice
introduction to efficient hinged dissection of polyregulars, and is also where
our research began.

\subsection{Small Polyominoes}

\begin{figure}
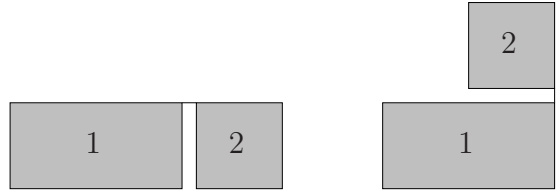

\centerline{\input tromino_2.pstex_t}
\caption{\label{trominoes}
  Two-piece hinged dissection of all trominoes.}
\end{figure}

\begin{figure}
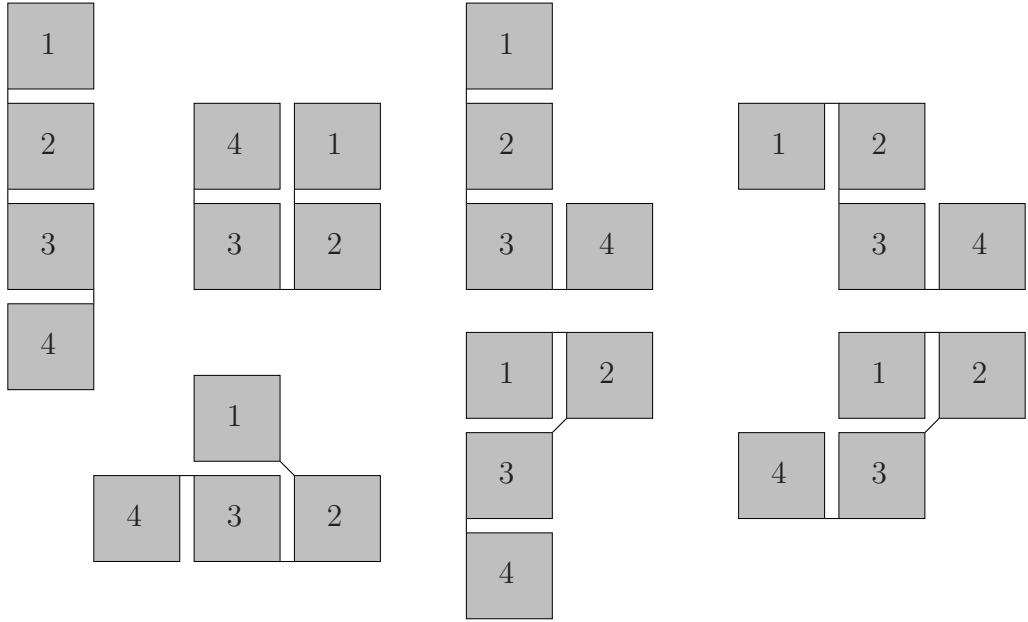

\centerline{\input tetromino_4.pstex_t}
\caption{\label{tetrominoes}
  Four-piece hinged dissection rotated into all tetrominoes, while keeping
  the orientation of square 2 fixed.}
\end{figure}

Constructing a hinged dissection that forms any $n$-omino is easy for small
values of $n$.  There is only one monomino and one domino, so no hinges are
necessary.  There are two trominoes, and a two-piece dissection is easy to
find; see Figure \ref{trominoes}.  The natural four-square dissection can
be hinged to rotate into all tetrominoes; see Figure \ref{tetrominoes}.

Unfortunately, in contrast to normal dissections, dividing an $n$-omino into
its constituent squares is insufficient for it to hinge into all other
$n$-ominoes for $n = 5$:

\begin{theorem} \label{pentomino lower bound}
Five identical squares cannot be hinged in such a way that they can be rotated
into all pentominoes.
\end{theorem}

\begin{proof}
Suppose there were a hinging $H$ of five squares that could rotate into every
pentomino.  We first consider the I-pentomino, which is a $1 \times 5$
rectangle.  Because $H$ rotates into the I-pentomino, the squares must be
hinged one after the other in a chain, ordered by their position in the
rectangle.  An example without the hinges is given in Figure \ref{I-pentomino}.

\begin{figure}
\centerline{%
  \begin{tabular}{ccccc}
    $\vcenter{\hbox{\input{pentomino_lb_I.pstex_t}}}$ & \hfil &
    $\vcenter{\hbox{\input{pentomino_lb_X.pstex_t}}}$ & \hfil &
    $\vcenter{\hbox{\input{pentomino_lb_T.pstex_t}}}$ \\
    \parbox{1.6in}{
      \caption{\label{I-pentomino} I-pentomino without hinges}
    } & &
    \parbox{1.7in}{
      \caption{\label{X-pentomino} X-pentomino with sample hinging}
    } & &
    \parbox{1.7in}{
      \caption{\label{T-pentomino} T-pentomino with forced hinges}
    }
  \end{tabular}
}
\end{figure}

We next consider the X-pentomino, in which the five squares form a (Greek)
cross.  Four of the five squares are arms of the X-pentomino.  Each of these
four has two adjacent vertices that do not touch any other square, and thus
cannot have hinges on them.  It follows that none of these four squares can
have hinges at diagonally opposite vertices.  Thus, at most one of the five
squares can have hinges at diagonally opposite vertices.
Figure \ref{X-pentomino} gives an example of such a hinging of the X-pentomino
(there are several).  In this example, only square 3, the middle square, has
hinges at diagonally opposite vertices.

We next consider the T-pentomino, in which three squares are stacked one on top
of the other, and the other two squares are to the right and to the left of the
top square in the stack.  One end of the chain (call it square 1) must be on
the bottom of the stack, because it is adjacent to only one other square (which
must necessarily be square 2).  The top middle square cannot be square 3, for
otherwise it would be impossible to connect all the squares in a chain.  Thus,
in particular, the other end of the chain (square 5) must be either the left or
the right square at the top.  This argument limits us to the configuration in
Figure \ref{T-pentomino} and its mirror image.



Suppose that we transform $H$ from the T-pentomino to the I-pentomino while
leaving square 2 in the same orientation.  If the I-pentomino lies
horizontally, then square 1 must rotate $180^\circ$ clockwise, causing square 2
to have hinges at diagonally opposite vertices.
Square 3 must rotate
$90^\circ$ clockwise, square 4 rotates $270^\circ$ clockwise, and square 5
rotates $90^\circ$ clockwise, as shown in Figure \ref{T to I horizontal}.  But
this requires that two squares, squares 2 and 4, have hinges at diagonally
opposite vertices.  Because this possibility has been ruled out, we cannot
transform $H$ from the T-pentomino to the I-pentomino while leaving square 2 in
the same orientation and having the I-pentomino lie horizontally.

Suppose that we transform $H$ from the T-pentomino to the I-pentomino while
leaving square 2 in the same orientation, with the I-pentomino standing
vertically.  Then square 3 must rotate $90^\circ$ counterclockwise.  This would
leave both of its hinges adjacent to square 2, as shown in Figure~\ref{T to I
vertical}.  Clearly squares 3 and 4 cannot be connected
in this way.

This exhausts all cases for transforming $H$ from the T-pentomino to the
I-pentomino.  Thus the desired hinging $H$ does not exist.
\end{proof}

\begin{figure}
\centerline{%
  \begin{tabular}{ccc}
    $\vcenter{\hbox{\input{pentomino_lb_horiz.pstex_t}}}$ & \hfil &
    $\vcenter{\hbox{\input{pentomino_lb_vert.pstex_t}}}$ \\
    \parbox{2.5in}{
      \caption{\label{T to I horizontal} First try: T to I}
    } & &
    \parbox{2.5in}{
      \caption{\label{T to I vertical} Second try: T to I}
    }
  \end{tabular}
}
\end{figure}

\subsection{General Polyominoes}

\begin{figure}
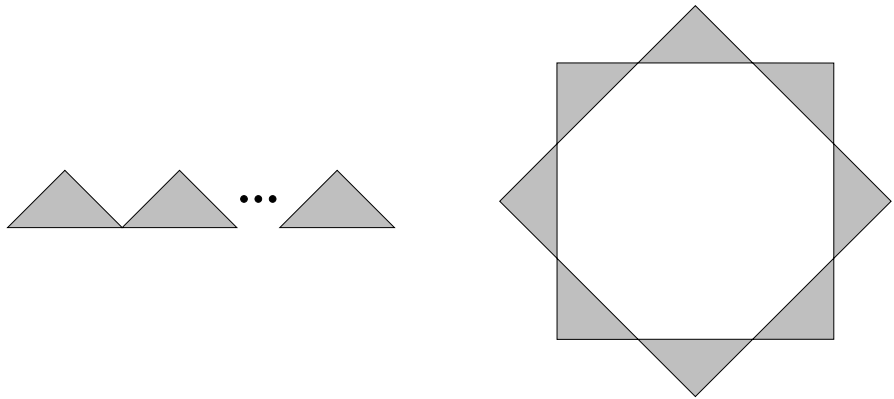

\centerline{\input polyomino_2n.pstex_t}
\caption{\label{polyomino 2n} $2n$-piece hinged dissection of all $n$-ominoes.
  (Left) Connected in a path.  (Right) Connected in a cycle, $n = 4$.}
\end{figure}

Our first hinged dissection of general $n$-ominoes uses $2n$ right isosceles
triangles; see Figure \ref{polyomino 2n}.  Note that a \emph{cyclicly hinged}
dissection (in which the pieces are connected in a cycle) is a stronger result
than a \emph{linearly hinged} dissection (in which the pieces are connected in
a path): simply breaking one of the hinges in the cycle results in a linearly
hinged dissection.

\begin{theorem} \label{theorem polyomino 2n}
A cycle of $2n$ right isosceles triangles, joined at their base vertices,
can be rotated into any $n$-omino.
\end{theorem}

\begin{figure}
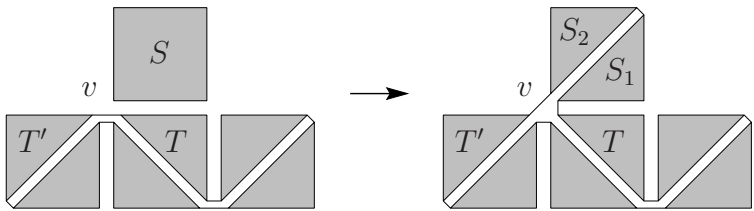

\centerline{\input adding_polyomino_2n.pstex_t}
\caption{\label{2n induction step}
  Attaching a square $S$ to the $2n$-piece hinged dissection of $n$-ominoes.}
\end{figure}

\begin{proof}
Apply Lemma \ref{build up} with $c=1$.  The case $n=1$ is shown in Figure
\ref{2-piece monomino}.

\begin{figure}
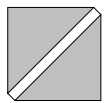

\centerline{\input monomino_2.pstex_t}
\caption{\label{2-piece monomino}
  A 2-piece hinged dissection of a monomino.}
\end{figure}

We can attach a square $S$ to the hinging of a polyomino $P$ as follows; refer
to Figure \ref{2n induction step}.  Let $T$ be the triangle in this hinging
that shares an edge with $S$.  One of its base vertices, say $v$, is also
incident to $S$, and it must be hinged to some other triangle $T'$.  We split
$S$ into two right isosceles triangles $S_1$ and $S_2$ so that both have a base
vertex at $v$.  Now we replace $T$'s hinge at $v$ with a hinge to $S_1$, and
add a hinge from $S_2$ to $T'$ at $v$.  Finally, $S_1$ and $S_2$ are hinged
together at their other base vertex.  The result is a hinging of $2n$ triangles
rotated into $P$.  We can optionally swap $S_1$ and $S_2$ in order to avoid
crossings between the hinges.
\end{proof}

Now we explain how to modify this dissection to use two fewer pieces:

\begin{corollary} \label{corollary polyomino 2n-2}
For $n \geq 2$, the $(2n-2)$-piece hinged dissection in Figure \ref{polyomino
2n-2} can be rotated into any $n$-omino.
\end{corollary}

\begin{figure}
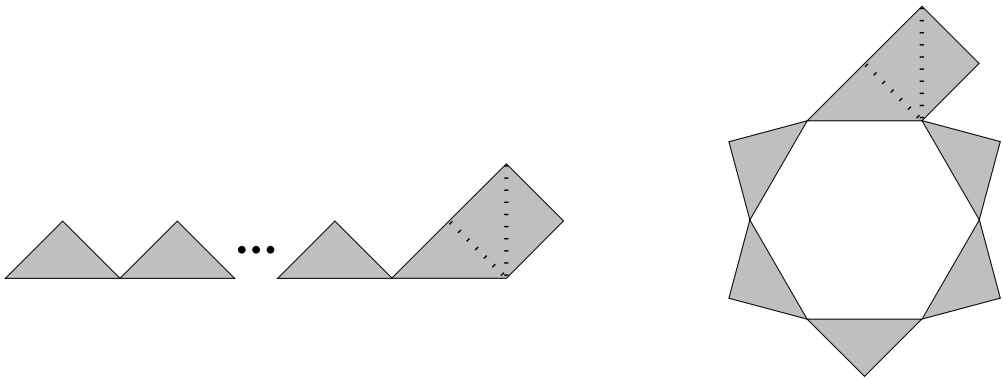

\centerline{\input polyomino_2nm2.pstex_t}
\caption{\label{polyomino 2n-2} $(2n-2)$-piece hinged dissection of all
  $n$-ominoes.  (Left) Linear.  (Right) Cyclic, $n=4$.}
\end{figure}

\begin{figure}
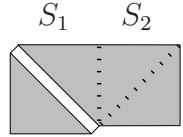

\centerline{\input domino_2.pstex_t}
\caption{\label{domino 2}
  A 2-piece hinged dissection of a domino.}
\end{figure}

\begin{proof}
Apply Lemma \ref{build up} with $c=2$.  The case $n=2$ is shown in Figure
\ref{domino 2}.  One square in this domino, $S_1$, has hinges at diagonally
opposite vertices just as before, but the other square, $S_2$, has only one
hinge.  By symmetry, we can arrange in Lemma \ref{build up} for $S_2$ to be
chosen as slippery, and hence all attachments act as in
Theorem \ref{theorem polyomino 2n}.
\end{proof}

This method will be generalized in Section \ref{Improved Polyregulars} to
support all polyregulars.

\section{Polyregulars}
\label{Polyregulars}

This section describes two methods for constructing, given any $n \geq 1$
and $k \geq 3$, a hinged dissection that can be rotated into all $n \times
k$-regulars.  In particular, this result generalizes the polyomino case
($k = 4$) of the previous section.
However, our first solution will not be as efficient as that in the
previous section; it serves as a simpler warm-up for the second solution.
The second solution combines the first solution and the efficient $k=4$
solution to obtain a single method that is efficient for all~$k$.

\subsection{Inefficient Polyregulars}
\label{Inefficient Polyregulars}

\begin{figure}
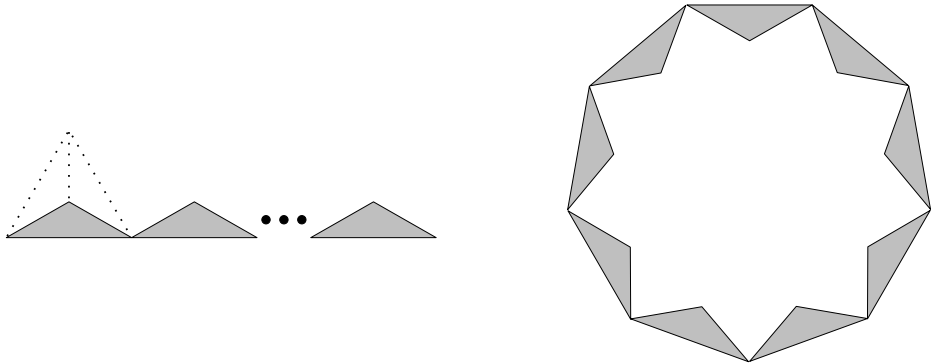

\centerline{\input polyiamond_3n.pstex_t}
\caption{\label{polyiamond 3n} $3n$-piece hinged dissection of all
  $n$-iamonds.  (Left) Linear.  (Right) Cyclic, $n = 3$.}
\end{figure}

Our first hinged dissection splits each regular $k$-gon into $k$ isosceles
triangles, by adding an edge from each vertex to the center of the polygon.
See Figure \ref{polyiamond 3n}.  Because it is rather difficult to draw a
generic regular $k$-gon, our figures will concentrate on the case of $k = 3$,
i.e., $n$-iamonds.  We define the \emph{base} of each isosceles triangle to be
the edge that coincides with an edge of the regular $k$-gon (whose length
differs from all the others unless $k=6$).  The \emph{base vertices} are the
endpoints of the base, and the \emph{opposite angle} is the interior angle of
the remaining vertex.

\begin{theorem} \label{theorem polyregular kn}
A cycle of $k n$ isosceles triangles with opposite angle $2\pi/k$, joined at
their base vertices, can be rotated into any $n \times k$-regular.
\end{theorem}

\begin{figure}
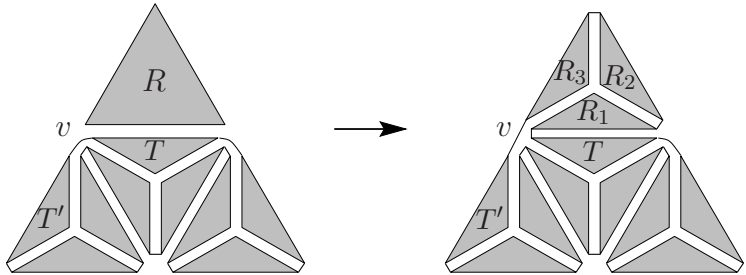

\centerline{\input adding_polyiamond_3n.pstex_t}
\caption{\label{3n induction step}
  Attaching an equilateral triangle $R$ to the $3n$-piece hinged dissection of
  $n$-iamonds.}
\end{figure}

\begin{proof}
Apply Lemma \ref{build up} with $c=1$.  The case $n=1$ is shown in Figure
\ref{3-piece moniamond}.

\begin{figure}
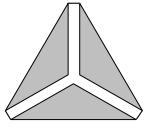

\centerline{\input moniamond_3.pstex_t}
\caption{\label{3-piece moniamond}
  A 3-piece hinged dissection of a moniamond.}
\end{figure}

We can attach a regular $k$-gon $R$ to the hinging of an $n \times k$-regular
$P$ as follows; refer to Figure \ref{3n induction step}.  Let $T$ be the
triangle in this hinging that shares an edge with $R$.  Both of its base
vertices are also incident to $R$.  Let $v$ be either one of $T$'s base
vertices, and suppose that it is hinged to triangle $T'$.  We split $R$ into
$k$ isosceles triangles $R_1$, \dots, $R_k$ so that both $R_1$ and $R_k$ have a
base vertex at $v$.  Now we replace $T$'s hinge at $v$ with a hinge to $R_1$,
and add a hinge from $R_k$ to $T'$ at $v$.  Finally, $R_i$ and $R_{i+1}$ are
hinged together at their common base vertex, for all $1 \leq i < k$.  The
result is a hinging of $k n$ triangles rotated into $P$.  We can optionally
renumber $R_1$, \dots, $R_k$ as $R_k$, \dots, $R_1$ in order to avoid crossings
between the hinges.
\end{proof}

While the number of pieces will be improved dramatically in the next section,
we show that the trick of merging the last few pieces also applies to this
dissection, reducing the number of pieces by $k$:

\begin{corollary} \label{corollary polyregular kn-k}
For $n \geq 2$, the $k(n-1)$-piece hinged dissection in Figure \ref{polyiamond
3n-3} can be rotated into any $n \times k$-regular.
\end{corollary}

\begin{figure}
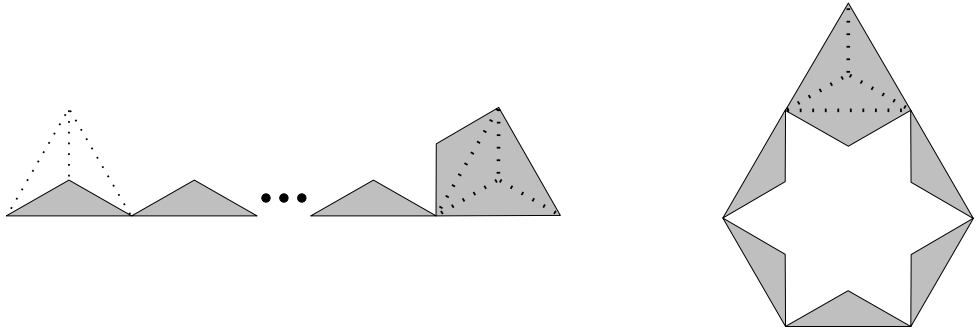

\centerline{\input polyiamond_3nm3.pstex_t}
\caption{\label{polyiamond 3n-3} $(3n-3)$-piece hinged dissection of all
  $n$-iamonds.  (Left) Linear.  (Right) Cyclic, $n=3$.}
\end{figure}

\begin{figure}
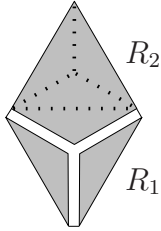

\centerline{\input diamond_3.pstex_t}
\caption{\label{diamond 3}
  A 3-piece hinged dissection of a diamond.}
\end{figure}

\begin{proof}
Apply Lemma \ref{build up} with $c=2$.  The case $n=2$ is shown in Figure
\ref{diamond 3}.  One regular $k$-gon $R_1$ in this $2 \times k$-regular has
hinges at all its vertices just as before, but the other regular $k$-gon $R_2$
has only two hinges.
By symmetry, we can arrange in Lemma \ref{build up} for $R_2$ to be
chosen as slippery, and hence all attachments act as in
Theorem \ref{theorem polyregular kn}.
\end{proof}

\subsection{Improved Polyregulars}
\label{Improved Polyregulars}

The goal of this section is to improve the previous hinged dissection
for polyregulars so that, for $k=4$, the number of pieces matches
the method in Section \ref{Polyominoes} for polyominoes.  To see how to do
this, let us compare the two methods when restricted to $k=4$.  The method in
Section \ref{Inefficient Polyregulars} splits each square into four right
isosceles triangles; i.e., it makes four cuts to the center of the square.
In contrast, the method in Section \ref{Polyominoes} makes only two of these
cuts.  In other words, the method in Section \ref{Polyominoes} can be thought
of as merging adjacent pairs of right isosceles triangles from the method in
Section \ref{Inefficient Polyregulars}.

This discussion suggests the following generalized improvement to the method
in Section \ref{Inefficient Polyregulars}, for arbitrary $k$:
join adjacent pairs of right isosceles
triangles, until zero or one triangles are left.  For even $k$ (like $k=4$),
this will halve the number of pieces; and for odd $k$, it will almost halve the
number of pieces.  The intuition behind why this method will work is that when
we added a regular $k$-gon to an existing polyregular in the proof of Theorem
\ref{theorem polyregular kn}, we had two existing hinges at which we could
connect the new $k$-gon; at most halving the number of hinges will still leave
at least one hinge to connect the new $k$-gon.

\begin{figure}
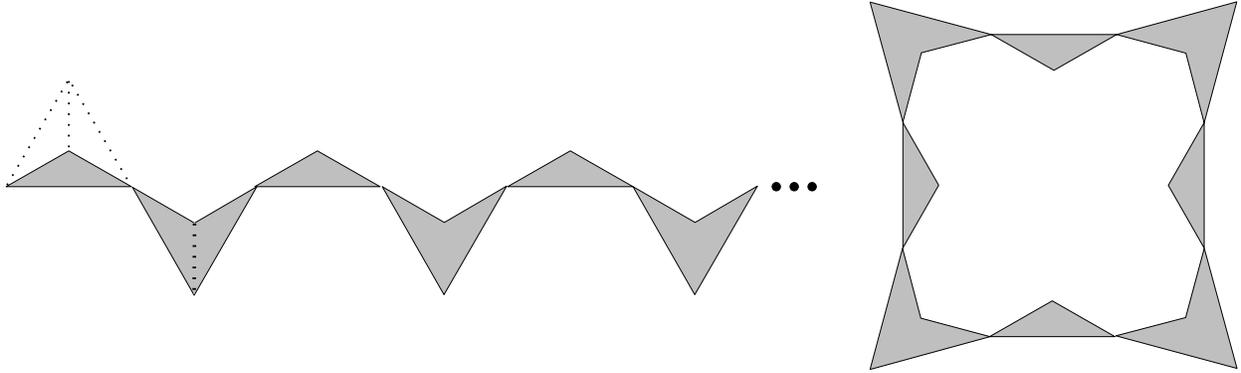

\centerline{\input polyiamond_2n.pstex_t}
\caption{\label{polyiamond 2n}
  $2n$-piece hinged dissection of all $n$-iamonds.
  (Left) Linear.  (Right) Cyclic, $n=4$.}
\end{figure}

In general, our hinged dissection will consist of $\lceil k/2 \rceil n$ pieces.
If $k$ is even, every piece will be the union of two isosceles triangles, each
with opposite angle $2\pi/k$, joined along an edge other than the base.  If $k$
is odd, every group of $\lfloor k/2 \rfloor$ of these pieces is followed by a
single isosceles triangle with opposite angle $2 \pi/k$.  For example, for
polyiamonds ($k=3$), the pieces alternate between single triangles and
``double'' triangles (see Figure \ref{polyiamond 2n}).  Independent of
the parity of $k$, the pieces are joined at the base vertices of the
constituent triangles that have not been merged to other base vertices.

Again, our figures will focus on the case $k=3$, as in Figure \ref{polyiamond
2n}.

\begin{theorem} \label{theorem polyregular kd2n}
The described cyclicly hinged dissection of $\lceil k/2 \rceil n$ pieces
can be rotated into any $n \times k$-regular.
\end{theorem}

\begin{figure}
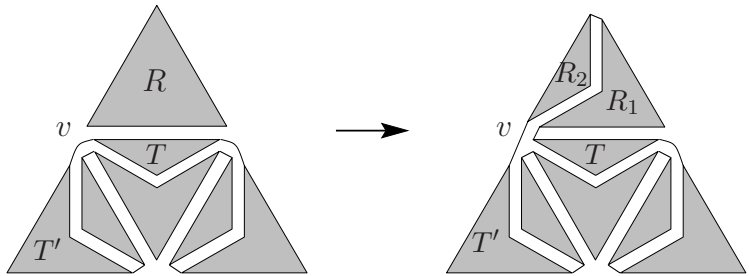

\centerline{\input adding_polyiamond_2n.pstex_t}
\caption{\label{kd2n induction step}
  Attaching an equilateral triangle $R$ to the $2n$-piece hinged dissection of
  $n$-iamonds.}
\end{figure}

\begin{proof}
Apply Lemma \ref{build up} with $c=1$.  The case $n=1$ is shown in Figure
\ref{2-piece moniamond}.

\begin{figure}
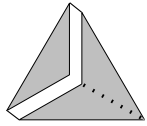

\centerline{\input moniamond_2.pstex_t}
\caption{\label{2-piece moniamond}
  A 2-piece hinged dissection of a moniamond.}
\end{figure}

We can attach a regular $k$-gon $R$ to the hinging of an $n \times k$-regular
$P$ similar to the proof of Theorem \ref{theorem polyregular kn}; refer to
Figure \ref{kd2n induction step}.  Let $T$ be the piece in this hinging that
shares an edge with $R$.  Both of the base vertices of one of its constituent
triangles are incident to $R$.  Let $v$ be such a base vertex of $T$ that is
not joined to a base vertex of another constituent triangle of $T$ (which we
call \emph{lone} base vertices), and suppose that it is hinged to piece $T'$.
We split $R$ into $\lceil k/2 \rceil$ pieces $R_1$, \dots, $R_{\lceil k/2
\rceil}$ so that both $R_1$ and $R_{\lceil k/2 \rceil}$ have a lone base vertex
at $v$.  Now we replace $T$'s hinge at $v$ with a hinge to $R_1$, and add a
hinge from $R_{\lceil k/2 \rceil}$ to $T'$ at $v$.  We hinge $R_i$ and
$R_{i+1}$ together at their common lone base vertex, for all $1 \leq i < \lceil
k/2 \rceil$.  Finally, if $k$ is odd, we choose one of the pieces $R_1$, \dots,
$R_{\lceil k/2 \rceil}$ to be a single triangle instead of a double triangle,
appropriately so that the single triangles appear periodically
in the resulting cycle of pieces, with a period of $\lceil k/2 \rceil$.
The result is the desired hinging of $\lceil k/2 \rceil n$ pieces
rotated into $P$.  We can optionally renumber $R_1$, \dots,
$R_{\lceil k/2 \rceil}$ as $R_{\lceil k/2 \rceil}$, \dots, $R_1$ in order to
avoid crossings between the hinges.
\end{proof}

Our final hinged dissection of polyregulars improves the previous one by
$\lceil k/2 \rceil$ pieces.

\begin{corollary} \label{corollary polyregular kd2tnm1}
For $n \geq 2$, the $\lceil k/2 \rceil (n-1)$-piece hinged dissection in
Figure \ref{kd2nm1 hinging} can be rotated into any $n \times k$-regular.
\end{corollary}

\begin{figure}
\centerline{\input polyiamond_2nm2.pstex_t}
\caption{\label{kd2nm1 hinging}
  $(2n-2)$-piece hinged dissection of all $n$-iamonds.
  (Left) Linear.  (Right) Cyclic, $n=4$.}
\end{figure}

\begin{figure}
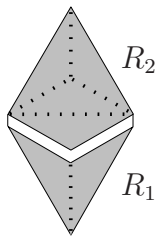

\centerline{\input diamond_2.pstex_t}
\caption{\label{diamond 2}
  A 2-piece hinged dissection of a diamond.}
\end{figure}

\begin{proof}
Apply Lemma \ref{build up} with $c=2$.  The case $n=2$ is shown in Figure
\ref{diamond 2}.  One regular $k$-gon $R_1$ in this $2 \times k$-regular has
hinges at roughly half of its vertices just as before, but the other regular
$k$-gon $R_2$ has only two hinges.
By symmetry, we can arrange in Lemma \ref{build up} for $R_2$ to be
chosen as slippery, and hence all attachments act as in
Theorem \ref{theorem polyregular kd2n}.
\end{proof}

\section{Polyabolos}
\label{Polyabolos}

Another well-studied class of polyforms that does not fall under the class of
polyregulars is \emph{polyabolos}, the union of equal-size half-squares (right
isosceles triangles) joined at equal-length edges.  In this section,
we present a hinged dissection of polyabolos.

\begin{figure}
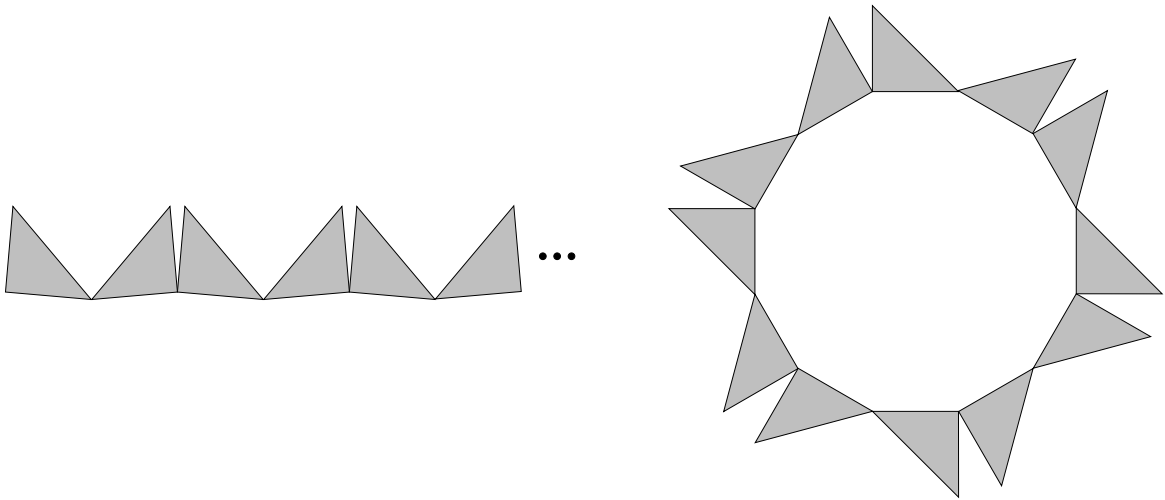

\centerline{\input polyabolo_4n.pstex_t}
\caption{\label{polyabolo 4n}
  $4n$-piece hinged dissection of all $n$-abolos.
  (Left) Linear.  (Right) Cyclic, $n = 3$.}
\end{figure}

Our dissection is a cycle of $4n$ right isosceles triangles, as shown in
Figure \ref{polyabolo 4n}.  Like Figure \ref{polyomino 2n}, the triangles
point outward, but unlike Figure \ref{polyomino 2n}, they are joined at a short
edge instead of the long edge.  The orientations of the triangles (or
equivalently, which of the two short edges we connect to the
other triangles) alternate along the cycle.

\begin{theorem} \label{theorem polyabolo 4n}
The $4n$-piece hinged dissection in Figure \ref{polyabolo 4n} can be rotated
into any $n$-abolo.
\end{theorem}

\begin{figure}
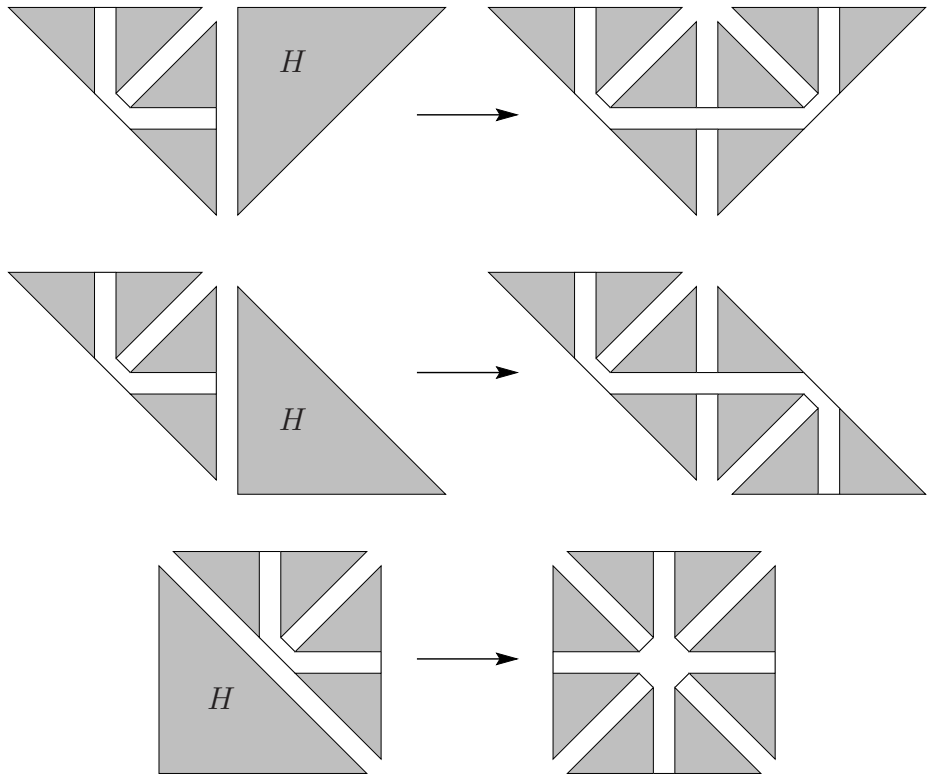

\centerline{\input adding_polyabolo_4n.pstex_t}
\caption{\label{4n induction step}
  Attaching a half-square $H$ to the $4n$-piece hinged dissection of $n$-abolos.}
\end{figure}

\begin{proof}
Apply Lemma \ref{build up} with $c=1$.  The case $n=1$ is shown in Figure
\ref{4-piece monabolo}.  Note that in contrast to all previous dissections,
there are no hinges at the vertices of the monabolo.  There are, however,
hinges at the midpoints of all the edges.

\begin{figure}
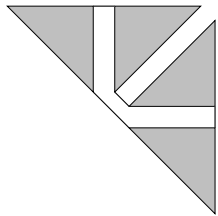

\centerline{\input monabolo_4.pstex_t}
\caption{\label{4-piece monabolo}
  A 4-piece hinged dissection of a monabolo.}
\end{figure}

We can attach a half-square $H$ to the hinging of a polyabolo $P$ at these
midpoints as shown in Figure \ref{4n induction step}.  There are three cases
according to relative orientations of $H$ and the incident half-square.  But in
all cases we obtain the same hinged dissection (Figure \ref{polyabolo 4n}) with
triangles pointing outward from the cycle, and alternating in orientation along
the cycle.
\end{proof}

\label{Different-Size Polyominoes}

An interesting consequence of this theorem is a hinged dissection that
can be rotated into polyominoes with squares of different sizes:

\begin{corollary}
A common hinged dissection can be rotated into any $n$-omino and any
$2n$-omino.
\end{corollary}

\begin{figure}
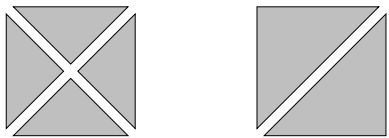

\centerline{\input n_2n_ominoes.pstex_t}
\caption{\label{n 2n ominoes}
  Two ways to split a polyomino into a polyabolo.}
\end{figure}

\begin{proof}
Both can be viewed as a $4n$-abolo, by splitting a square in the $n$-omino
into four pieces (Figure \ref{n 2n ominoes}, left) and splitting a square in
the $2n$-omino into two pieces (Figure \ref{n 2n ominoes}, right).
\end{proof}

This dissection uses a large number of pieces, namely $16 n$.  In fact, we can
do much better by simply using the $4n$-piece path dissection of the
$2n$-omino from Figure \ref{polyomino 2n}, left.
On the one hand, as in Theorem \ref{theorem polyomino 2n},
the hinged dissection can rotate into any $2n$-omino.
On the other hand, we can rotate the pieces and view adjacent
pairs of pieces as (temporarily) merged along their short sides,
and we obtain a hinged dissection that can rotate into any $n$-omino.
The full cyclic hinging does not work, because the $2n$-omino wants the long
sides inside while the $n$-omino wants them outside,
so rotating from one to the other would require twisting the hinges.

\section{Other Polyforms}
\label{Other Polyforms}

An interesting open problem is whether there is a hinged dissection that can be
rotated into all polyforms of a particular size and type.  In other words, for
a fixed $n$ and polygon $P$, is there a hinged dissection that rotates into any
connected edge-to-edge gluing of $n$ copies of $P$?  As a step towards solving
this problem, we present a hinged dissection for a large class of polyforms of
type $P$.  Specifically, we impose the restriction that for any two copies of
$P$ sharing an edge, there must be a rigid motion (a combination of
translations and rotations) that

  \begin{enumerate}
  \item takes one copy of $P$ to the other copy, and
  \item takes the shared edge in one copy to the shared edge in
        the other copy.
  \end{enumerate}

Such a polyform is called a \emph{restricted polyform}.  The first constraint
says that the copies of $P$ are not flipped over.
The second constraint says that only ``corresponding edges''
of copies of $P$ are joined.  This is actually not that uncommon: if $P$ is
generic in the sense that no two edges have the same length, then the second
constraint is implied by the edge-to-edge condition.

Comparing to our previous results, every polyregular is a polyform satisfying
the described restriction.  However, polyabolos do not satisfy the second
constraint; for example, if we join two right isosceles triangles so that their
union forms a larger right isosceles triangle, then noncorresponding edges
have been joined.

\begin{figure}
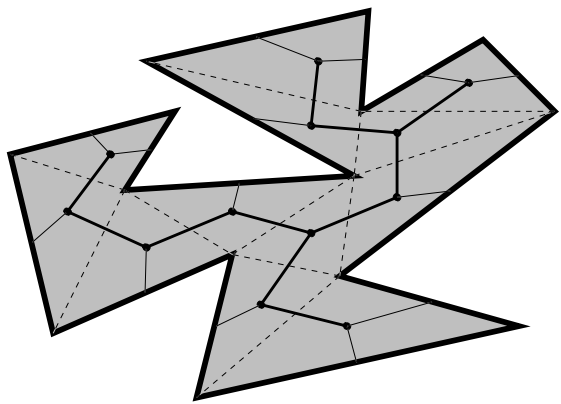

\centerline{\input cut_up.pstex_t}
\caption{\label{cut up}
  Cutting up a polygon $P$ with cuts through the midpoint of every edge.
  Only the dashed lines are not cuts; they show the underlying triangulation
  $T$.  The thick solid lines and dots form the dual tree $D$,
  and the thin solid lines are the cuts from dual vertices to edge midpoints.}
\end{figure}

The method for hinge-dissecting restricted polyforms works as follows.  We
subdivide $P$ by making cuts incident to the midpoint of every boundary edge,
so that there is one piece surrounding each vertex of $P$.  This can be done as
follows; refer to Figure \ref{cut up}.  Take a triangulation $T$ of $P$.
First, cut along edges of the triangulation so that the remaining connections
between triangles form a dual tree $D$; this step adds artificial ``edges''
to $P$ to make $P$ simply connected (hole-free).
Second, position each vertex of $D$ anywhere interior to the corresponding
triangle in $T$, and cut along the edges of~$D$.  Third, cut from each
vertex of $D$ to the midpoint of every edge of $P$ incident to the
corresponding triangle of~$T$.

\begin{figure}
\centerline{\input{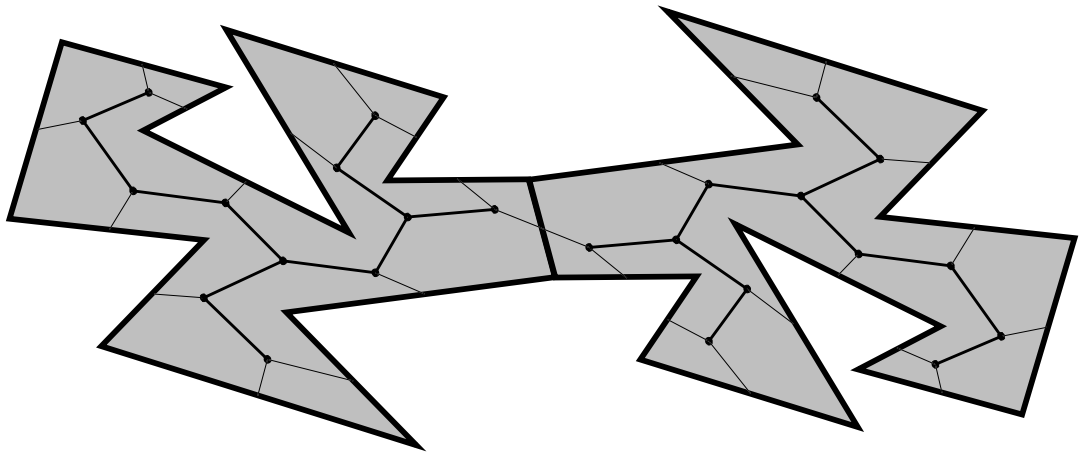}}
\caption{\label{cut up join}
  Joining two copies of $P$, once each is cut up: switching over from one copy
  of $P$ to the other does not affect the order of shapes of pieces we visit.}
\end{figure}

Now the actual hinged dissection is simple: repeat the cyclic decomposition of
$P$, $n$ times, and hinge the pieces at the midpoints of the edges of $P$.  Now
at any edge of $P$ we can decide to visit an incident copy of $P$ before
completing the traversal of $P$, and we visit the same sequence of pieces.  See
Figure \ref{cut up join} for a simple example.

In this way we can construct any restricted polyform of type $P$, nearly
proving the following theorem:

\begin{theorem} \label{restricted polyforms}
There is a $k n$-piece cyclicly hinged dissection that can be rotated into any
restricted $n$-form of type $P$, where $P$ is a polygon with $k$ vertices.
\end{theorem}

\begin{proof}
There is one detail omitted in the discussion above, so let us go through a
formal proof.  Apply Lemma \ref{build up} with $c=1$.  (While this lemma was
designed for general $n$-forms, it applies equally well to restricted
$n$-forms.)  The case $n=1$ just takes the decomposition of $P$ described
above, and hinges it at the midpoints of edges of $P$.  Adding a copy of $P$ to
a restricted polyform of type $P$ requires special care.  Let $Q$ denote the
copy of $P$ to which we want to attach $P$, and let $e$ denote the edge of $Q$
to which $P$ will attach.  We need to place $P$ against $e$ such that the rigid
motion mapping $Q$ to $P$ and $e$ to $e$ also maps the pieces of $Q$ to the
pieces of $P$ (where ``pieces'' refer to the subdivision described above).

Certainly there is a rigid motion $m$ mapping the pieces of $Q$ to the pieces
of $P$, so we should attach $P$'s edge $m(e)$ to $Q$'s edge $e$.  The only
possible wrinkle is that if $P$ has symmetry, in the sense that there is a
rigid motion $s$ from $P$ to itself, then we might instead attempt to attach
$s(m(e))$ to $e$.  Fortunately, we get to choose the orientation of the
subdivision of $P$ for the copy we are attaching.  We can explore all symmetric
versions of the subdivision of $P$ and choose the one that places $m(e)$
against $e$.
\end{proof}

\section{Polyforms of Different Types}
\label{Polyforms of Different Types}

\subsection{Different Restricted Polyforms}

Figure \ref{Dudeney} shows that a regular $3$-gon can be hinge-dissected into a
regular $4$-gon.  This example suggests the more general possibility of
hinge-dissecting $n \times 3$-regulars into $n \times 4$-regulars.  Indeed,
such a hinged dissection is possible, even for restricted polyforms of
arbitrary types:

\begin{theorem} \label{k-way polyforms}
Given a hinged dissection $H$ that rotates into polygons
$P_1, P_2, \dots, P_k$,
there is a cyclicly hinged dissection $H'$ that rotates into all restricted
$n$-forms of type $P_i$ for all $1 \leq i \leq k$.  If $H$ has $n$ pieces and
$P_i$ has $p_i$ sides, then $H'$ has at most $3n-3+\sum_{i=1}^k p_i$ pieces.
\end{theorem}

Before we discuss applications of this theorem, let us prove it.  First we need
a result which is interesting in its own right: any hinged dissection can be
turned into a cyclicly hinged dissection.
By removing one hinge from the cyclicly hinged dissection,
we can also obtain a linearly hinged dissection.

\begin{lemma} \label{lemma chain to cyclic}
For any dissection $H$, there is a cyclicly hinged dissection $H'$ that can be
rotated into any polygon that $H$ can.
If $H$ has $n$ pieces, then $H'$ has at most $3n-3$ pieces.
\end{lemma}

\begin{proof}
The graph of the hinging structure (in which vertices represent pieces and
edges represent hinges) can be any planar graph.  First take a spanning tree of
that graph, and remove all other hinges.  This transformation removes the
cycles from the hinging structure while keeping the pieces connected.
Now for each (original) piece with only one hinge, we cut it along a polygonal
line from that hinge to any other point on the boundary (e.g., another vertex),
and add a hinge between the two pieces at that point.  For each original piece
that has at least two hinges, we cut along a tree of line segments that is
interior to the piece and has leaves at the hinges, and we replace each
original hinge with two ``parallel'' hinges.
The result is a cyclicly hinged dissection $H'$,
which can be rotated as $H$ can because it simply subdivides and adds more
degrees of motion.  See Figure \ref{chain to cyclic} for an example.

\begin{figure}
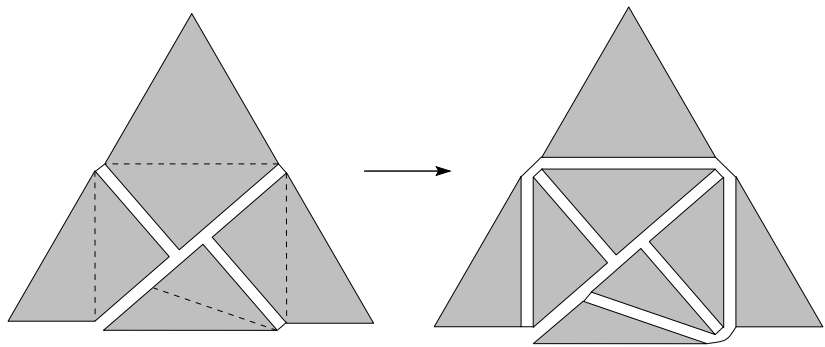

\centerline{\input tri2sqr_cyclic.pstex_t}
\caption{\label{chain to cyclic}
  Converting the linearly hinged dissection in Figure \protect\ref{Dudeney}
  into a cyclicly hinged dissection.}
\end{figure}

If $H$ has $n$ pieces, then when we reduce the hinging structure to a spanning
tree it has $n-1$ hinges.  Our construction doubles every original hinge, and
adds an additional hinge for every leaf (a piece adjacent to only one original
hinge), for a total of at most $3n-3$.  There is one piece per hinge in a
cyclicly hinged dissection, so the number of pieces in $H'$ is at most $3n-3$.
\end{proof}

We are now in the position to show how to hinge-dissect polyforms of different
types.

\medskip
\smallskip

\begin{proof}[Theorem \ref{k-way polyforms}]
Start with the cyclicly hinged dissection from Lemma \ref{lemma chain to
cyclic}.  Now we want to add cuts so that there is a hinge at the midpoint of
each edge in $P_i$ for all $i$.  This can be done as follows.
For each $i \in \{1, \dots, k\}$, and for each edge $e$ of $P_i$ that does not
already have a hinge at its midpoint, consider the piece $Q$ whose boundary
contains $e$'s midpoint when $H$ is rotated into $P_i$.
Refer to Figure \ref{tri2sqr one step}.
Let $q_1$ and $q_2$ denote the paths of $Q$'s boundary connecting the two
hinges incident to $Q$, where the paths include their endpoints.
Order $q_1$ and $q_2$ so that the midpoint of $e$ is on $q_1$.
Pick an arbitrary point $r$ on $q_2$,
add a polygonal cut from $r$ to the midpoint of~$e$, and add a hinge connecting
the two pieces at the midpoint of $e$.
Figure \ref{tri2sqr one step} shows the special case in which $r$ is chosen
to be an endpoint of $q_2$, i.e., a hinge.
In this case, the hinge at $r$ is assigned to the piece of $Q$ that is not
incident to the other endpoint of $q_2$, and the hinged dissection remains
cyclic.

\begin{figure}
\centerline{\input{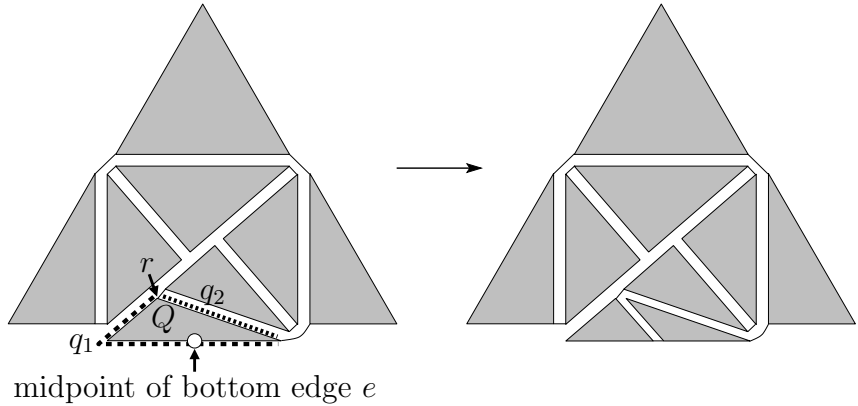}}
\caption{\label{tri2sqr one step}
  In the cyclicly hinged dissection from Figure \protect\ref{chain to cyclic},
  cutting piece $Q$ into two pieces so that the midpoint of the triangle's
  bottom edge $e$ becomes a hinge.}
\end{figure}

Performing this operation for all choices of $i$ and $e$, we obtain a cyclicly
hinged dissection $\hat H$ that can be repeated $n$ times to obtain $H'$.
See Figure \ref{inefficient polyiamond to polyomino} for a complete example.
Each repetition of $\hat H$ can be thought of, in particular,
as a subdivision of $P_i$ with hinges at the midpoints.  Thus, as we proved in
Theorem \ref{restricted polyforms}, $H'$ can be rotated into any restricted
polyform of type $P_i$, and this holds for any $i$.

\begin{figure}
\centerline{\input{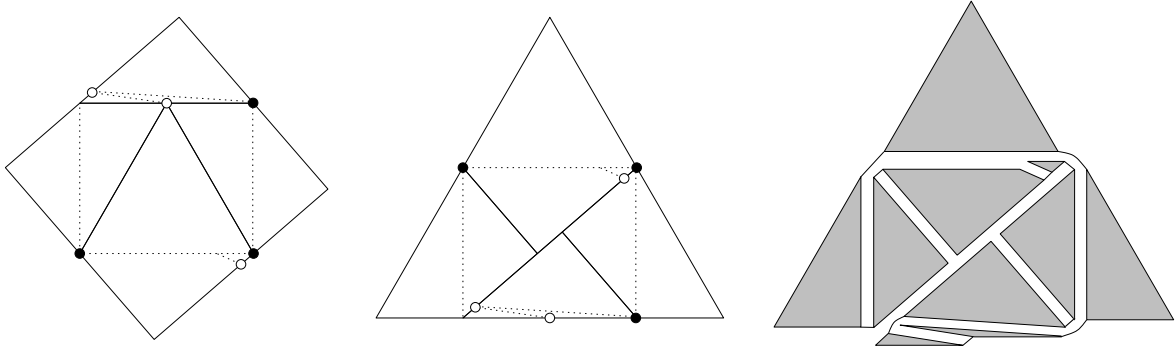}}
\caption{\label{inefficient polyiamond to polyomino}
  Converting the cyclicly hinged dissection in Figure \protect\ref{chain to
  cyclic} (roughly)
  into one with hinges at the midpoints when rotated into either shape.
  Filled circles are hinges; open circles are midpoints from both shapes.}
\end{figure}

We started with the $(3n{-}3)$-piece cyclicly hinged dissection from Lemma
\ref{lemma chain to cyclic}, and added at most one piece per midpoint of an
edge of a polygon $P_i$.  Therefore, we added $\sum_{i=1}^k p_i$ pieces, for
a total of $3n-3+\sum_{i=1}^k p_i$.
\end{proof}

\subsection{$n \times k$-Regulars to $n \times k'$-Regulars}

One particularly interesting application of the previous result is that,
provided there is a hinged dissection of a regular $k$-gon into a regular
$k'$-gon, there is a hinged dissection for each $n \geq 1$ that rotates into
all $n \times k$-regulars and all $n \times k'$-regulars.
In this section, we explore more efficient hinged dissections for
polyiamonds, polyominoes, and polyhexes.

To this end, we will use a more powerful technique for building such a hinged
dissection.  A linearly hinged dissection $H$ between regular polygons $P$ and
$Q$ is called \emph{extendible} if
  \begin{enumerate}
  \item When $H$ is rotated into $P$, every edge of $P$ has a hinge at an
        endpoint or at its midpoint.
  \item When $H$ is rotated into $Q$, every edge of $Q$ has a hinge at an
        endpoint or at its midpoint.
  \item There is a vertex in the last piece of the chain 
        and a vertex in the first piece of the chain
        such that these vertices coincide
        when $H$ is rotated into $P$ and when $H$ is rotated into $Q$.
  \end{enumerate}
\noindent
The new flexibility, which will allow us to use fewer pieces in our
dissections, is for pieces to be connected by a common vertex instead of just
by midpoints.  This variation works only because $P$ and $Q$ are regular
polygons, and so we can exploit their symmetries.
Note also that there is a subtle difference between having a cyclicly hinged
dissection and having a chain whose ends coincide: we need the ability to
``invert'' the chain, which would require twisting a hinge if the two ends
were joined together by a hinge.


\begin{theorem} \label{theorem extendible chains}
If there is an extendible linearly hinged dissection $C$
between a regular $k$-gon and a regular $k'$-gon,
then the chain $C^n$ that is formed by concatenating
$n$ copies of $C$ rotates into any $n \times k$-regular and into any $n \times
k'$-regular, for all $n \geq 1$.
\end{theorem}

\begin{proof}
Apply Lemma \ref{build up} with $c=1$, for both $k$ and $k'$ symmetrically; we
will focus on $k$.  The case $n=1$ is solved by the given hinged dissection
$C$.  Now consider attaching a regular $k$-gon $P$ to a rotation of $C^{n-1}$
into some $n \times k$-regular $R$.  Let $P'$ be the polygon to which we are
attaching $P$.  Let $p$ be a hinge at the midpoint or an endpoint of the edge
of $P'$ to which we are attaching $P$, which is guaranteed to exist because $C$
is extendible.  Split $C^{n-1}$ into two chains, $C_p^{n-1,1}$ and
$C_p^{n-1,2}$ at $p$.


Point $p$ is either the midpoint or an endpoint.  If it is the midpoint, make
cuts in $P$ such that the cuts in $P \cup P'$ exhibit $180^\circ$-rotational
symmetry about $p$.
Split $C$ at point $p$ into $C^1_p$ and $C^2_p$, and
splice them together at the original endpoints to give $C_p$.  Then splice
$C_p$ into $C_p^{n-1,1}$ and $C_p^{n-1,2}$, and the result is $C^n$ rotated
into the desired polyregular $R \cup P$.

If $p$ is an endpoint of the common side of $P$ and $P'$, make cuts in $P$ so
that $P \cup P'$ are identical with respect to $\alpha$-rotational symmetry
about $p$, where $\alpha$ is the interior angle of $P$.
We can then form $C^n$ by cutting and splicing as
in the previous case.  Again $C^n$ rotates into the desired polyregular $R \cup
P$.
\end{proof}



\begin{figure}
\centerline{\includegraphics{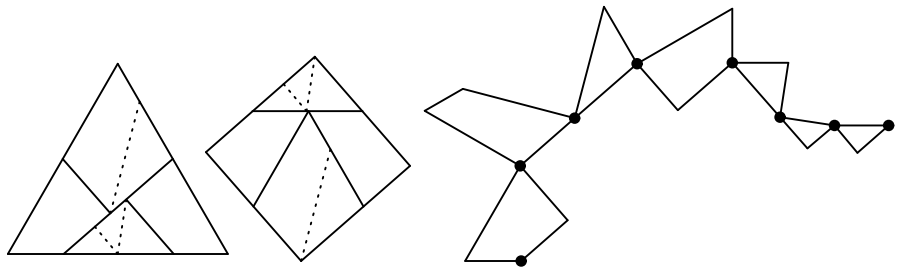}}
\caption{\label{efficient triangle to square}
  Triangle to a square.  (Left) Dissection.  (Right) Extendible chain.}
\end{figure}

Our goal now is to find efficient extendible linearly hinged dissections
between equilateral triangles, squares, and regular hexagons.  We start with
the first two shapes, by modifying the hinged dissection in Figure
\ref{Dudeney}.  We add three more cuts:
  \begin{enumerate}
  \item from the midpoint of the base of the equilateral triangle
        to the right angle in the small triangular piece,
  \item from the midpoint of the base of the equilateral triangle
        to a point (say, the midpoint) on the left leg of the small triangular
        piece, and
  \item from the right angle of the largest piece to a point near the apex
        of the equilateral triangle.
  \end{enumerate}
These additional cuts (shown dashed) produce the extendible linearly hinged
dissection in Figure~\ref{efficient triangle to square}.

\begin{corollary}
For any positive $n$, there is a linearly hinged dissection of $7n$ pieces that
rotates into all $n$-iamonds and all $n$-ominoes.
\end{corollary}

\begin{proof}
This follows directly from Theorem \ref{theorem extendible chains} and the
above construction.
\end{proof}

\begin{figure}
\centerline{\includegraphics{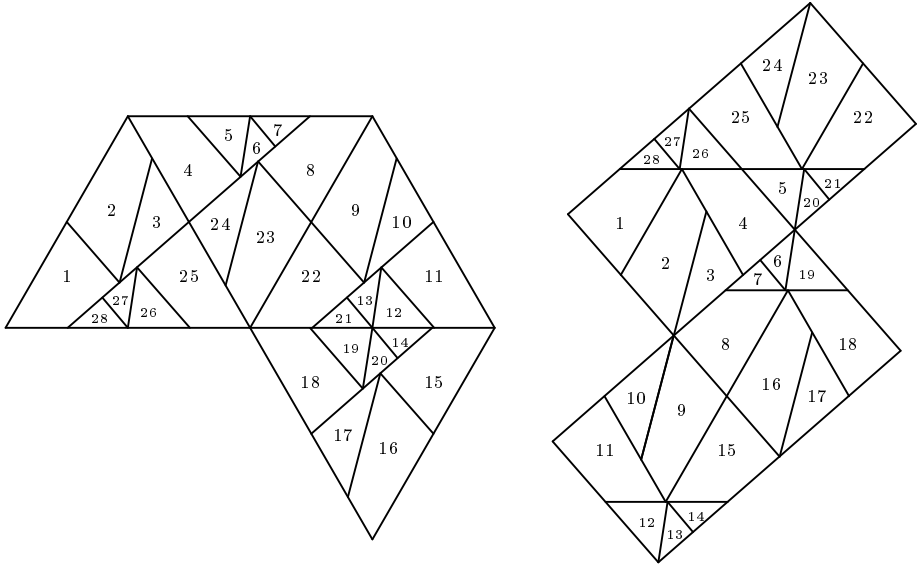}}
\caption{\label{tetriamond and tetromino}
  Rotating the chain in Figure \protect\ref{efficient triangle to
  square}, repeated four times, into a tetriamond (left)
  and tetromino (right).}
\end{figure}

As an example, we show how to form a tetriamond and a tetromino in Figure
\ref{tetriamond and tetromino}.  There are $7n = 28$ pieces in the chain $C^4$.
We number these pieces in order from $1$ to $28$.  Note that a piece repeats in
$C^4$ after six other pieces, so that, for example, pieces $3$, $10$, $17$, and
$24$ are identical.

Analogous to the situation in Corollary \ref{corollary polyomino 2n-2} and
Figure \ref{polyomino 2n-2}, we can save on the number of pieces by judicious
merging.  With respect to the particular tetriamond and tetromino shown, we can
merge pieces $22$ through $28$ together.  For any $n>2$, we can always merge
the last seven pieces together.  A proof of this, however, requires more care
in the ordering of the last several squares (and triangles) chosen in the
inductive proof.


Let us next consider $n$-ominoes and $n$-hexes.  There are several five-piece
dissections of a regular hexagon to a square \cite{Lindgren-1972,
Frederickson-1997}, but the best that is known for hinged dissections has six
pieces \cite{Frederickson-2002}.  One such dissection is linearly hinged, but
adapting it to make it extendible in a few number of additional pieces seems
difficult.

\begin{figure}
\centerline{\includegraphics{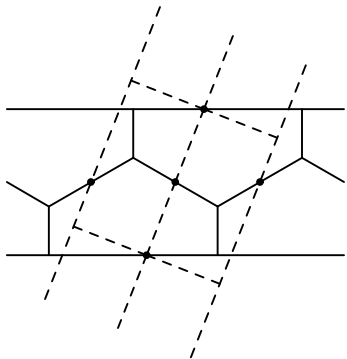}}
\caption{\label{strips hexagon to square}
  Crossposed strips for a hexagon to a square.}
\end{figure}

Thus we start afresh, deriving a TT2-strip dissection by crossposing strips as
in Figure \ref{strips hexagon to square}.  (See \cite{Frederickson-1997} for a
discussion of the T-strip technique.)  The hexagon strip consists of halves of
hexagons, cut from the midpoint of one side to the midpoint of the opposite
side.  The square strip consists of rectangles; each is half of a square.  The
boundaries of the rectangles cross the sides of the hexagons at their
midpoints, indicated by the dots.

As discussed in \cite{Frederickson-2002}, crossing T-strips at midpoints gives
rise to hinge points.  Other hinge points result from cutting the hexagon and
square in half at the midpoints of sides, and placing these halves in the strip
so that the resulting vertices touch.  This dissection was inspired by an
analogous 8-piece hinged dissection of a hexagon to a Greek cross in
\cite{Frederickson-2002}.

\begin{figure}
\centerline{\includegraphics{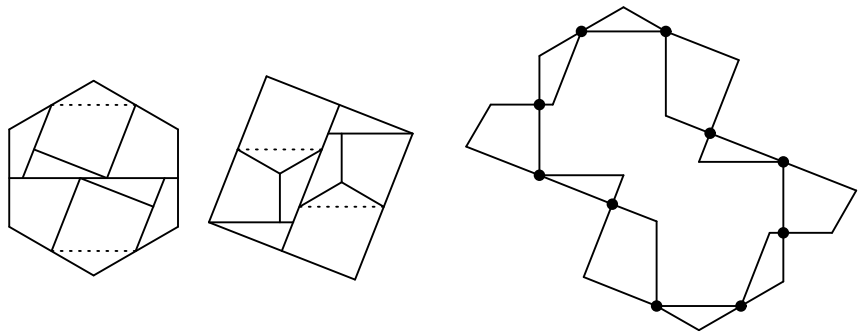}}
\caption{\label{efficient hexagon to square}
  Hexagon to a square. (Left) Dissection. (Right) Extendible cycle.}
\end{figure}

The dissection derived from the crossposition in Figure \ref{strips hexagon to
square} is cyclicly hingeable but does not have hinges on all six sides of the
hexagon.  We thus add two additional cuts (shown dashed) to produce the
dissection on the left of Figure~\ref{efficient hexagon to square}.  It is
cyclicly hinged as shown on the right.  Splitting the cycle at any hinge point
gives an extendible chain.

\begin{corollary}
For any positive $n$, there is a linearly hinged dissection of $10 n$ pieces
that rotates into all $n$-ominoes and $n$-hexes.
\end{corollary}

\begin{proof}
Again this follows directly from Theorem \ref{theorem extendible chains} and
the above construction.
\end{proof}

We have not studied how many pieces can be merged together
to save a few pieces in the case that $n>1$.

\begin{figure}
\centerline{\includegraphics{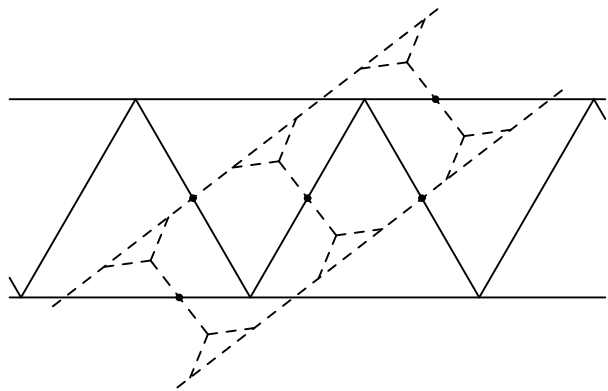}}
\caption{\label{strips triangle to hexagon}
  Crossposed strips for an equilateral triangle to a regular hexagon.}
\end{figure}

For handling $n$-iamonds and $n$-hexes, we crosspose two strips as in Figure
\ref{strips triangle to hexagon}.  The hexagon strip is created by slicing two
isosceles triangles from the hexagon, with each slice going through the
midpoints of two of the hexagon's sides.  These midpoints are identified with
dots.  The appropriate angle between the crossposed strips is found by forcing
the midpoints of the remaining sides to be positioned on sides of the
equilateral triangle.
This gives a 6-piece hinged dissection, matching the
fewest pieces known for any hinged dissection
of an equilateral triangle to a hexagon \cite{Frederickson-2002}.

\begin{figure}
\centerline{\includegraphics{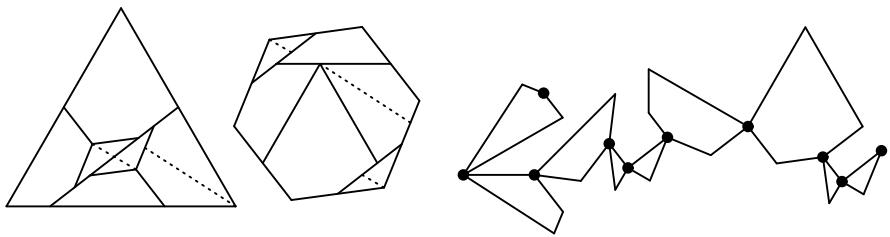}}
\caption{\label{efficient triangle to hexagon}
  Triangle to hexagon.  (Left) Dissection.  (Right) Extendible chain.}
\end{figure}

To get an extendible linearly hinged dissection, we make three more cuts (shown
dashed), giving the dissection in Figure~\ref{efficient triangle to hexagon}.
Two of the cuts are through the isosceles triangles, producing a linearly
hinged dissection.  The third cut is to a vertex of the equilateral triangle,
to put a hinge at the base and right side of the equilateral triangle.

\begin{corollary}
For any positive $n$, there is a linearly hinged dissection of $9n$ pieces that
rotates into all $n$-iamonds and $n$-hexes.
\end{corollary}

Again, this corollary follows directly from Theorem \ref{theorem extendible
chains}, and we have not studied how many pieces can be merged together to save
a few pieces in the case that $n>1$.

%

\section{Conclusion}
\label{Conclusion}

Our most general result is that, for any hinged dissection $H$ and $n \geq 1$,
there is a hinged dissection $H'$ that rotates into any arrangement of $n$
copies of $P$ joined at corresponding edges, where $P$ is any polygon into
which $H$ rotates.  In particular, if $H$ is a single-piece ``dissection,''
there is a hinged dissection that rotates into all arrangements of copies of a
given polygon $P$ joined at corresponding edges.  This statement includes
polyregulars as a subclass, for which we showed how to improve the number of
pieces.  This class contains as subclasses several well studied objects:
polyominoes, polyiamonds, and polyhexes.
We proved the analogous result for polyabolos
(equal-size right isosceles triangles joined edge-to-edge), which do not fall
under any of the above classes, but are still considered ``polyforms.''  Using
more general (multipiece) dissections for $H$, we showed how to simultaneously
hinge-dissect all polyiamonds and polyominoes; all polyiamonds and polyhexes;
and all polyominoes and polyhexes---in general, $n \times k$-regulars and
$n \times k'$-regulars when there is a hinged dissection of a regular $k$-gon
into a regular $k'$-gon.

Following up on our work, Frederickson \cite[pp.~234--236]{Frederickson-2002}
has shown how to obtain similar results for \emph{twist-hinge} dissections,
in which hinges cannot be rotated but can be twisted
(flipped over $180^\circ$).  In particular, for $n \geq 5$,
he describes a $(4n-5)$-piece linearly twist-hinged dissection that twists
into all $n$-ominoes.  He also describes, for $n \geq 4$, a $6n$-piece
linearly twist-hinged dissection that twists into all $n$-iamonds, and when $n$
is even, into all $(n/2)$-hexes.


Let us conclude with a list of interesting open problems about hinged
dissections, focusing on polyforms:

\begin{enumerate}
\item Can our results be generalized to arbitrary polyforms, that is,
  connected edge-to-edge gluings of $n$ nonoverlapping copies of a common
  polygon?
\item How many pieces are needed for a hinged dissection of all pentominoes?
  What about general $n$-ominoes as a function of $n$?  We know of no
  nontrivial lower bounds.
\item Can any $n$-omino be hinge-dissected into any $m$-omino (of an
  appropriate scale), for all $n, m$?  In Section \ref{Different-Size
  Polyominoes}, we proved this is true for $m = 2n$.
\item Can any regular $k$-gon be hinge-dissected into any regular $k'$-gon
  with the same area?
  Hinged dissections for ten different pairs of regular polygons appear in
  \cite{Frederickson-2002}.
\item Is there a single hinged dissection of all
  $n \times k_1$-regulars, $n \times k_2$-regulars, and
  $n \times k_3$-regulars?  For example, is there a hinged dissection
  that rotates into all $n$-iamonds, $n$-ominoes, and $n$-hexes?
  This question is equivalent to asking whether there is a hinged dissection
  that rotates into an equilateral triangle, square, and regular hexagon.
\item We have shown that there exist rotations of a common hinged
  dissection into any $n \times k$-regular.  Is it possible to
  continuously rotate the dissection from one configuration to another, while
  keeping the pieces nonoverlapping?  It is known that some hinged dissections
  cannot be continuously moved in this way \cite{Frederickson-2002}.
\item It would be interesting to generalize to higher dimensions.
  For example, \emph{polycubes} are connected face-to-face gluings of
  nonoverlapping unit (solid) cubes joined face-to-face.
  Can a collection of solids be hinged together at edges
  so that the dissection can be rotated into any $n$-cube (for fixed $n$)?
\end{enumerate}

\section*{Acknowledgments}

We thank Therese Biedl and Anna Lubiw for helpful discussions.
We also thank the anonymous referees for their comments.

\bibliography{dissect,polyominoes}
\bibliographystyle{plain}

\end{document}